\documentclass[cits,a4paper]{pos}

\usepackage{amsmath,amssymb}
\usepackage{subfigure}
\usepackage{graphicx}

\newcommand{\mps}{m_\mathrm{PS}}
\newcommand{\fps}{f_\mathrm{PS}}

\newcommand{\nft}{N_\mathrm{f}=2}
\newcommand{\preprint}{\newline%
  \begin{picture}(0,0)
  \put(300,120){\rm\small DESY 09-063,~~SFB/CPP-09-38}
  \end{picture}}

\graphicspath{{plots/}}

\title{$\chi$PT description of the pion mass and decay constant from
  $\nft$ twisted mass QCD\preprint}

\ShortTitle{$\chi$PT description of the pion mass and decay constant
  from $\nft$ tmQCD}

\author{Petros Dimopoulos, Roberto Frezzotti\\
  Dipartimento di Fisica, Universit\`a di Roma ``Tor Vergata''\\
  Via della Ricerca Scientifica 1, 00133 Rome, Italy\\
  E-mail: \email{\{dimopoulos,frezzotti\}@roma2.infn.it}}

\author{\speaker{Gregorio Herdoiza}, Karl Jansen\\
  NIC, DESY\\
  Platanenallee 6, 15738 Zeuthen, Germany\\
  E-mail: \email{\{Gregorio.Herdoiza,Karl.Jansen\}@desy.de}}

\author{Chris Michael\\
  Theoretical Physics Division, Department of Mathematical Sciences,
  University of Liverpool\\
  Liverpool L69 3BX, UK\\
  E-mail: \email{c.michael@liverpool.ac.uk}}

\author{Carsten Urbach\footnote{New affiliation: Helmholtz-Institut
    f{\"u}r Strahlen- und Kernphysik (Theorie) and Bethe Center for
    Theoretical Physics, Universit{\"a}t Bonn, 53115 Bonn,
    Germany}\\ Humboldt-Universit{\"a}t zu Berlin, Institut f{\"u}r
  Physik,\\ Newtonstr. 15, 12489 Berlin, Germany\\ E-mail:
  \email{Carsten.Urbach@physik.hu-berlin.de}}

\author{for the ETM Collaboration}

\abstract{We study lattice QCD determinations of the pion mass and
  decay constant by means of chiral perturbation theory ($\chi$PT).
  The lattice data are obtained from large scale simulations with
  $\nft$ flavours of twisted mass fermions at maximal twist. We
  perform a scaling test to the continuum limit of these data and use
  $\chi$PT to perform both the chiral and the infinite volume
  extrapolations.}

\FullConference{International Workshop on Effective Field Theories:
  from the pion to the upsilon \\ February 2-6 2009\\ Valencia, Spain}

\begin{document}

\section{Introduction}

In the last few years, lattice QCD simulations have made a substantial
progress in controlling the systematic effects present in the
determination of several important physical quantities allowing
therefore for their direct contact with experiment (see
\cite{Jansen:2008vs,Scholz:2009yz} for recent reviews). Simulations
including the light-quark flavours in the sea, as well as the strange
and recently also the charm, with pseudoscalar masses below $300$~MeV,
lattice extents $L > 2.0$~fm and lattice spacings smaller than
$0.1$~fm are currently being performed by several lattice groups. Such
simulations will eventually allow for an extrapolation of the lattice
data to the physical point and to the continuum limit while keeping
also the finite volume effects under control.

The European Twisted Mass collaboration (ETMC) has performed large
scale simulations with $\nft$ flavours of mass degenerate quarks using
Wilson twisted mass fermions at maximal twist. Four values of the
lattice spacing ranging from $0.1$~fm down to $0.051$~fm, pseudoscalar
masses between $280$ and $650$~MeV as well as several lattice sizes
($2.0 - 2.5$~fm) are used to address the systematic effects. 

The light pseudoscalar meson is an appropriate hadron for
investigating the systematic effects arising from continuum,
infinite volume and chiral extrapolations, because its mass and decay
constant can be obtained with high statistical precision in lattice
simulations.  Moreover, chiral perturbation theory ($\chi$PT) is best
understood for those two quantities. As a consequence of this study
one can extract other quantities of phenomenological interest, such as
the $u,d$ quark masses, the chiral condensate or the low energy
constants of $\chi$PT. First results for the pseudoscalar mass $\mps$
and decay constant $\fps$ from these $\nft$ simulations can be found
in Refs.~\cite{Boucaud:2007uk, Urbach:2007rt, Dimopoulos:2007qy,
  Boucaud:2008xu, Dimopoulos:2008sy,scalingnf2}.

ETMC is currently performing $N_{\rm f}=2+1+1$ simulations including
in the sea, in addition to the mass degenerate light $u,d$ quark
flavours, also the heavier strange and charm degrees of freedom. Some
first results for the pseudoscalar mass and decay constant from this
novel setup were presented in~\cite{Baron:2008xa,Baron:2009zq}.

In the following we will concentrate on the analysis of the $\nft$
data for $\mps$ and $\fps$.

\section{Lattice Action and Setup}

In the gauge sector we employ the tree-level Symanzik improved gauge
action (tlSym)~\cite{Weisz:1982zw}. The fermionic action for two
flavours of maximally twisted, mass degenerate quarks in the so-called
twisted basis~\cite{Frezzotti:2000nk,Frezzotti:2003ni} reads
\begin{equation}
  \label{eq:sf}
  S_\mathrm{tm}\ =\ a^4\sum_x\left\{ \bar\chi(x)\left[ D[U] + m_0 +
      i\mu_q\gamma_5\tau^3\right]\chi(x)\right\}\, ,
\end{equation}
where $m_0$ is the untwisted bare quark mass tuned to its critical
value $m_\mathrm{crit}$, $\mu_q$ is the bare twisted quark mass,
$\tau^3$ is the third Pauli matrix acting in flavour space and $D[U]$
is the Wilson-Dirac operator.

At maximal twist, i.e.~$m_0=m_\mathrm{crit}$, physical observables are
automatically O$(a)$ improved without the need to determine any action
or operator specific improvement coefficients~\cite{Frezzotti:2003ni}
(for a review see Ref.~\cite{Shindler:2007vp}). With this being the
main advantage, one drawback of maximally twisted mass fermions is
that flavour symmetry is broken explicitly at finite value of the
lattice spacing, which amounts to O$(a^2)$ effects in physical
observables.

For details on the setup, tuning to maximal twist and the analysis
methods we refer to Refs.~\cite{Boucaud:2007uk, Urbach:2007rt,
  Boucaud:2008xu}. Recent results for light quark masses, meson decay
constants, the pion form factor, $\pi$-$\pi$ scattering, the light
baryon spectrum, the $\eta'$ meson and the $\omega-\rho$ mesons mass
difference are available in Refs.~\cite{Blossier:2007vv,
  Frezzotti:2008dr, Feng:2009ij, Alexandrou:2008tn, Jansen:2008wv,
  McNeile:2009mx, Blossier:2009bx}.

Flavour breaking effects have been investigated for several
quantities~\cite{Boucaud:2007uk,Urbach:2007rt,Boucaud:2008xu,
  Dimopoulos:2008sy,Alexandrou:2008tn}. With the exception of the
splitting between the charged and neutral pion masses, other possible
splittings so far investigated are compatible with zero.  These
results are in agreement with a theoretical investigation using the
Symanzik effective
Lagrangian~\cite{Frezzotti:2007qv,Dimopoulos:2009qv}.

A list of the $\nft$ ensembles produced by ETMC can be found in
table~\ref{tab:setup}.
\begin{table}[t!]
  \centering
  \begin{tabular*}{0.85\textwidth}{@{\extracolsep{\fill}}ccrcccr}
    \hline\hline
    $\Bigl.\Bigr.$ Ensemble &$\beta$ & $a$~[fm] &$V/a^4$  
    & $\mps L$ & $a\mu_q$
    & $\mps$~[MeV]  \\ \hline\hline
    $D_1$ & $4.20$ & $ 0.051$ & $48^3 \cdot 96$ & $3.6$ & $0.0020$ & $280$ \\
    $D_2$ & & & $32^3 \cdot 64$ & $4.2$ & $0.0065$ & $510$  \\
    \hline
    $C_1$ & $4.05$ & $ 0.063$ & $32^3 \cdot 64$ & $3.3$ & $0.0030$ & $320$ \\
    $C_2$ & & & & $4.6$  & $0.0060$ & $450$ \\
    $C_3$ & & & & $5.3$ & $0.0080$ & $520$ \\
    $C_4$ & & & & $6.5$ & $0.0120$ & $630$ \\
    $C_5$ & & & $24^3 \cdot 48$& $3.5$ & $0.0060$ & $450$ \\
    $C_6$ & & & $20^3 \cdot 48$ & $3.0$ & $0.0060$ & $450$ \\
    \hline
    $B_1$ & $3.90$ & $ 0.079$ & $24^3 \cdot 48$ & $3.3$ & $0.0040$ & $330$ \\
    $B_2$ & & & & $4.0$ & $0.0064$ & $420$ \\
    $B_3$ & & & & $4.7$ & $0.0085$ & $480$ \\
    $B_4$ & & & & $5.0$ & $0.0100$ & $520$ \\
    $B_5$ & & & & $6.2$ & $0.0150$ & $640$ \\
    $B_6$ & & & $32^3 \cdot 64$ & $4.3$ & $0.0040$ & $330$ \\
    $B_7$ & & & & $3.7$ & $0.0030$ & $290$ \\
    \hline
    $A_2$ & $3.80$ & $ 0.100$ & $24^3 \cdot 48$ & $5.0$ & $0.0080$ & $ 410$ \\
    $A_3$ & & & & $5.8$ & $0.0110$ & $ 480$ \\
    $A_4$ & & & & $7.1$  & $0.0165$ & $ 580$ \\
    \hline\hline
  \end{tabular*}
  \caption{Ensembles with $\nft$ dynamical flavours produced by the
    ETM collaboration.  We give the ensemble name, the values of the
    inverse bare coupling $\beta=6/g_0^2$, an approximate value of the
    lattice spacing $a$, the lattice volume $V=L^3 \cdot T$ in lattice
    units, the approximate value of $\mps L$, the bare quark mass
    $\mu_q$ in lattice units and an approximate value of the light
    pseudoscalar mass $\mps$.}
  \label{tab:setup}
\end{table}

\section{Results}

\subsection{Scaling to the Continuum Limit}

Here we analyse the scaling to the continuum limit of the pseudoscalar
meson decay constant $\fps$ at fixed reference values of the
pseudoscalar meson mass $\mps$ and of the lattice size $L$ (we refer
to~\cite{Urbach:2007rt, Dimopoulos:2007qy, scalingnf2} for
details). The aim of this scaling test is to verify that
discretisation effects are indeed of $O(a^2)$ as expected for twisted
mass fermions at maximal twist.

In order to compare results at different values of the lattice spacing
it is convenient to measure the hadronic scale $r_0/a$
\cite{Sommer:1993ce}. It is defined via the force between static
quarks at intermediate distance and can be measured to high accuracy
in lattice QCD simulations. For details on how we measure $r_0/a$ we
refer to Ref.~\cite{Boucaud:2008xu}.

\begin{figure}[t]
  \centering
  \subfigure[\label{fig:fpsscaling}]%
  {\includegraphics[width=0.45\linewidth]{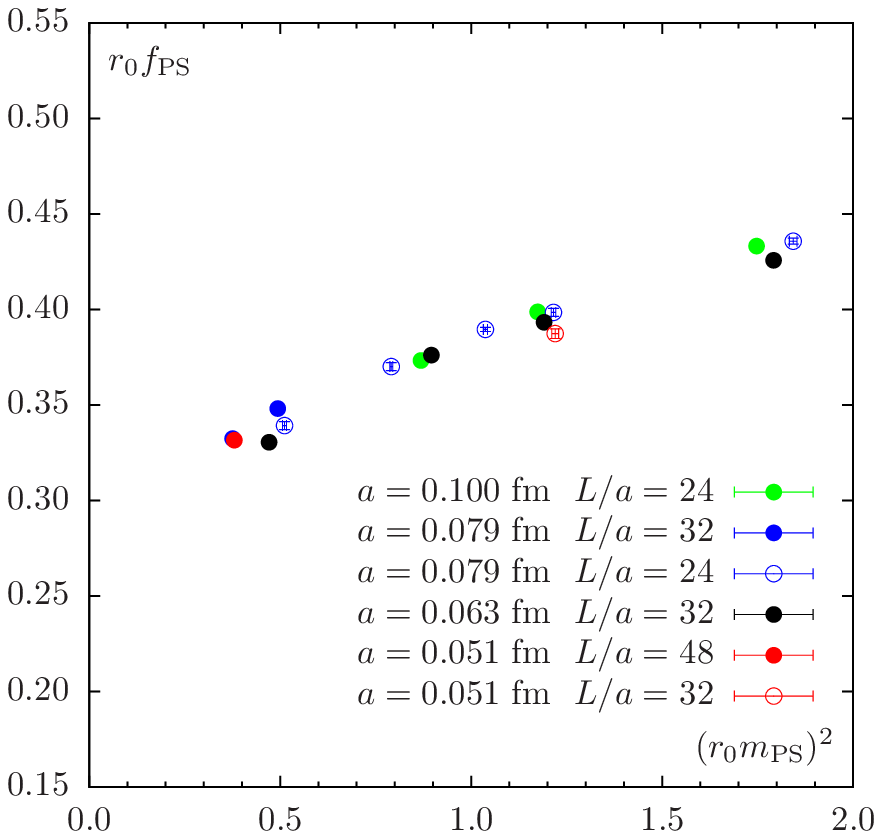}}
  \qquad
  \subfigure[\label{fig:scaling}]%
  {\includegraphics[width=0.455\linewidth]{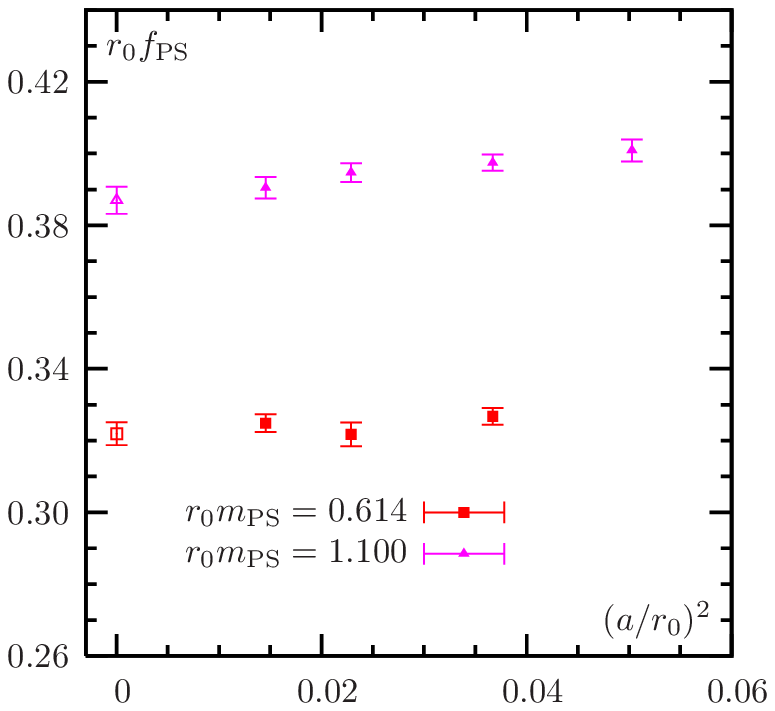}}
  \caption{(a) $r_0\fps$ as a function of $(r_0\mps)^2$ (b) Continuum
    limit scaling:~$r_0\fps$ as a function of $(a/r_0)^2$ at two fixed
    values of $r_0\mps$.}
  \label{fig:latart}
\end{figure}

In figure~\ref{fig:fpsscaling} we plot the results for $r_0\fps$ as a
function of $(r_0\mps)^2$. The vicinity of points coming from
different lattice spacings along a common curve is an evidence that
lattice artifacts are small for these quantities. This is indeed
confirmed in figure~\ref{fig:scaling} where the continuum scaling of
$r_0\fps$ is illustrated:~the very mild slope of the lattice data
shows that the expected O$(a^2)$ scaling violations are small. The
result of a linear extrapolation in $(a/r_0)^2$ to the continuum limit
is also shown.

\subsection{$\chi$PT Description of Finite Size Effects}

At the level of statistical accuracy we have achieved, finite size
effects (FSE) for $f_\mathrm{PS}$ and $m_\mathrm{PS}$ cannot be
neglected.  It is therefore of importance to study whether FSE can be
described within the framework of chiral perturbation theory. This
requires to compare simulations with different lattice volumes while
all other parameters are kept fixed, like for instance ensembles
$C_2$, $C_5$ and $C_6$ or $B_1$ and $B_6$ in
table~\ref{tab:setup}. For all these ensembles $m_\mathrm{PS}L\geq 3$
holds, which is believed to be needed for $\chi$PT formulae to
apply. Given the smallness of the lattice artifacts in $\fps$ and
$\mps$ (as discussed in the previous section), we proceed to compare
the \emph{measured} finite size effects to predictions of
\emph{continuum} $\chi$PT at NLO~\cite{Gasser:1986vb} (denoted GL) and
in the form of the resummed L{\"u}scher formula as described in
Ref.~\cite{Colangelo:2005gd} (for short CDH).

We note $R_O = [O(L=\infty)-O(L)]/O(L=\infty)$ the relative FSE for
the observable $O \in \{\mps, \fps\}$.  The results for
$R^\mathrm{meas}$, $R^\mathrm{GL}$ and $R^\mathrm{CDH}$ are compiled
in table~\ref{tab:R}. We observe that the CDH formulae tend to provide
an appropriate description of the lattice data. A more detailed
description of FSE in our $\fps$ and $\mps$ data was presented in
Ref.~\cite{Urbach:2007rt,scalingnf2}.

\begin{table}[t!]
  \centering
  \begin{tabular*}{0.95\textwidth}{@{\extracolsep{\fill}}lccr|rr}
    \hline\hline
    & $a$~[fm] & {\small$m_\mathrm{PS}L_1 \to m_\mathrm{PS}L_2$}
    & $R^{meas.}$[\%] 
    & $R^{GL}$   [\%] 
    & $R^{CDH}$  [\%] \\
    & & & [$L_1 \to L_2$] & [$L_1 \to L_2$] & [$L_1 \to L_2$] \\
    \hline\hline
    $m_\mathrm{PS}$ & $0.079$ & $3.3 \to 4.3$ & {$-1.8$} & {$-0.4$} &
        {$-1.0$} \\
        $f_\mathrm{PS}$ & $0.079$ & $3.3 \to 4.3$ & {$+2.6$} & {$+2.1$} &
        {$+2.3$} \\
        \hline
        $m_\mathrm{PS}$ & $0.063$ & $3.0 \to 4.6$ & $-6.1$  & $-1.7$ & $-5.9$\\
        $f_\mathrm{PS}$ & $0.063$ & $3.0 \to 4.6$ & $+10.7$ & $+6.3$ & $+8.5$\\
        \hline\hline
  \end{tabular*}
  \caption{Comparison of measured relative FSE, $R_O$, to estimates
    from $\chi$PT formulae.}
  \label{tab:R}
\end{table}

\subsection{$\chi$PT Description of the light-quark Mass Dependence}

The chiral extrapolation of lattice data down to the physical point is
currently one of the main sources of systematic uncertainties in the
lattice results. The possibility to rely on an effective theory such
as $\chi$PT to perform this extrapolation is therefore of great
importance to quote accurate results from lattice simulations. On the
other hand, while smaller quark masses are being simulated, the
possibility to perform a quantitative test of the effective theory as
well as to measure the low energy parameters of its Lagrangian becomes
more and more realistic.

We shall now present the results of a combined chiral, infinite volume
and continuum extrapolation of $\mps$ and $\fps$ for two values of the
lattice spacing (corresponding to $\beta=3.90$ and $\beta=4.05$). We
use $r_0/a$ to relate data from the two lattice spacings and a
non-perturbative determination of the renormalisation factor
$Z_\mathrm{P}$~\cite{Dimopoulos:2007fn} in order to perform the fit in
terms of renormalised quark masses. This analysis closely follows
those presented in Refs.~\cite{Urbach:2007rt, Dimopoulos:2007qy,
  Dimopoulos:2008sy,scalingnf2} to which we refer for more details.

We perform combined fits to our data for $\fps$, $\mps$, $r_0/a$ and
$Z_\mathrm{P}$ at the two values of $\beta$ with the formulae:
\begin{equation}
  \label{eq:fmps}
  \begin{split}
    r_0\fps &= r_0
    f_0\Bigl[1-2\xi\log\left(\frac{\chi_\mu}{\Lambda_4^2}\right) +
    T_f^\mathrm{NNLO} + D_{f_\mathrm{PS}}(a/r_0)^2\Bigr]\
    K_f^\mathrm{CDH}(L)\, , \\
    (r_0 \mps)^2 &= \chi_\mu r_0^2\Bigl[
    1+\xi\log\left(\frac{\chi_\mu}{\Lambda_3^2}\right)+
    T_m^\mathrm{NNLO} + D_{m_\mathrm{PS}}(a/r_0)^2\Bigr]\
    K_m^\mathrm{CDH}(L)^2\, ,\\
  \end{split}
\end{equation}
with $ \xi \equiv 2B_0\mu_R/(4\pi f_0)^2\ ,\,\chi_\mu\equiv
2B_0\mu_R\ ,\, \mu_R \equiv \mu_q/Z_\mathrm{P},\, f_0\equiv\sqrt{2}
F_0$.  $T_{m,f}^\mathrm{NNLO}$ denote the continuum NNLO terms of the
chiral expansion~\cite{Leutwyler:2000hx}, which depend on
$\Lambda_{1-4}$ and $k_M$ and $k_F$, and $K_{m,f}^\mathrm{CDH}(L)$ the
finite size corrections~\cite{Colangelo:2005gd}. Based on the form of
the Symanzik expansion in the small quark mass region, we parametrise
in eq.~(\ref{eq:fmps}) the leading cut-off effects by the two
coefficients $D_{f_{\rm PS},m_{\rm PS}}$. Setting $D_{f_{\rm
    PS},m_{\rm PS}}=0$ is equivalent to perform a constant continuum
extrapolation. Similarly, setting $T_{m,f}^\mathrm{NNLO}=0$
corresponds to fit to NLO $\chi$PT.

From the fit parameters coming from the quark mass dependence
predicted by $\chi$PT (in particular from $\Lambda_{3,4}$, $B_0$ and
$f_0$) the low energy constants $\bar\ell_{3,4}$ and the chiral
condensate $\Sigma$ can be determined.

By including or excluding data points for the heavier quark masses, it
is in principle possible to explore the regime of masses in which NLO
and/or NNLO SU$(2)~\chi$PT applies. We have actually generalised this
procedure in order to estimate all the dominant sources of systematic
uncertainties that can be addressed from our setup, which include,
discretisation effects, the order at which we work in $\chi$PT or
finite size effects. The idea is to use different fit ansatz (see
below) on a given data-set and to repeat this same procedure over
different data-sets: by weighting all these fits by their confidence
level we construct their distribution and estimate the systematic
error from the associated 68\% confidence interval.

The fit ans{\"a}tze we consider are:
\begin{itemize}
\item Fit A: NLO continuum $\chi$PT, $T_{m,f}^\mathrm{NNLO}\equiv0$,
  $D_{\mps,\fps}\equiv0$, priors for $r_0\Lambda_{1,2}$

\item Fit B: NLO continuum $\chi$PT, $T_{m,f}^\mathrm{NNLO}\equiv0$,
  $D_{\mps,\fps}$ fitted, priors for $r_0\Lambda_{1,2}$

\item Fit C: NNLO continuum $\chi$PT, $D_{\mps,\fps}\equiv0$,
  priors for $r_0\Lambda_{1,2}$ and $k_{M,F}$

\item Fit D: NNLO continuum $\chi$PT, $D_{\mps,\fps}$
  fitted, priors for $r_0\Lambda_{1,2}$ and $k_{M,F}$
\end{itemize}

The choice of the different data-sets (each of them including data for
different lattice spacings, quark masses and physical volumes) is made
in order to quantify how the quality of the fit is modified when
including/excluding data from {\it e.g.}, a given mass region or with
a given volume. The data-sets considered in the fits are listed in
Ref.~\cite{scalingnf2}.

\begin{figure}[t]
  \centering
  \subfigure[\label{fig:mps}]%
  {\includegraphics[width=0.45\linewidth]{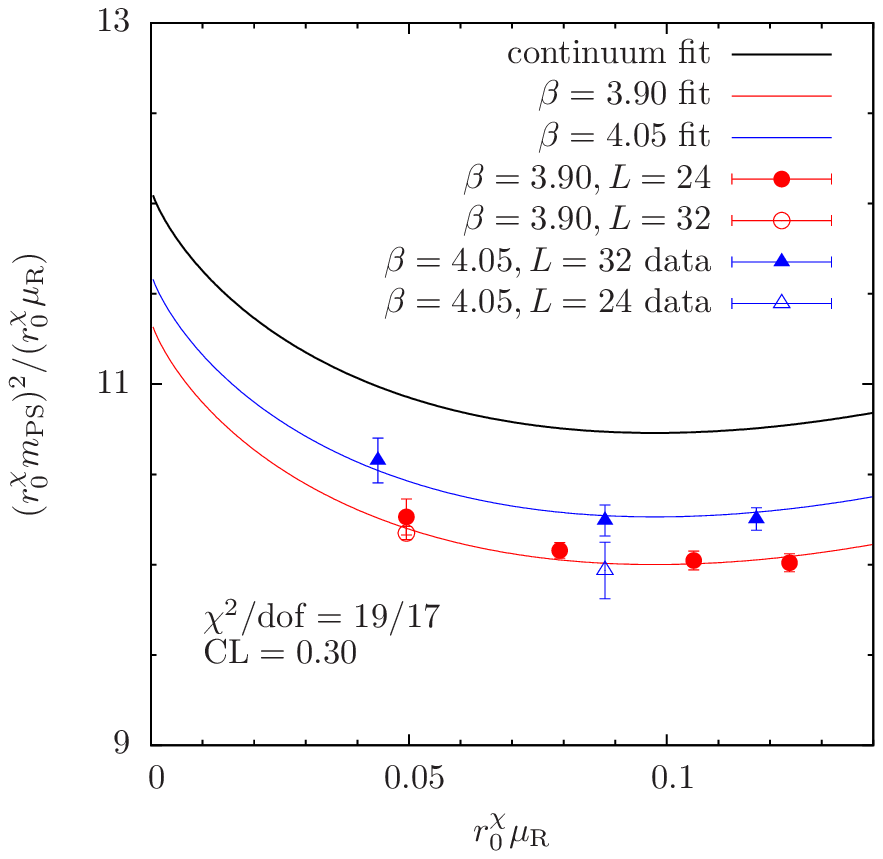}}
  \quad
  \subfigure[\label{fig:fpsasq}]%
  {\includegraphics[width=0.45\linewidth]{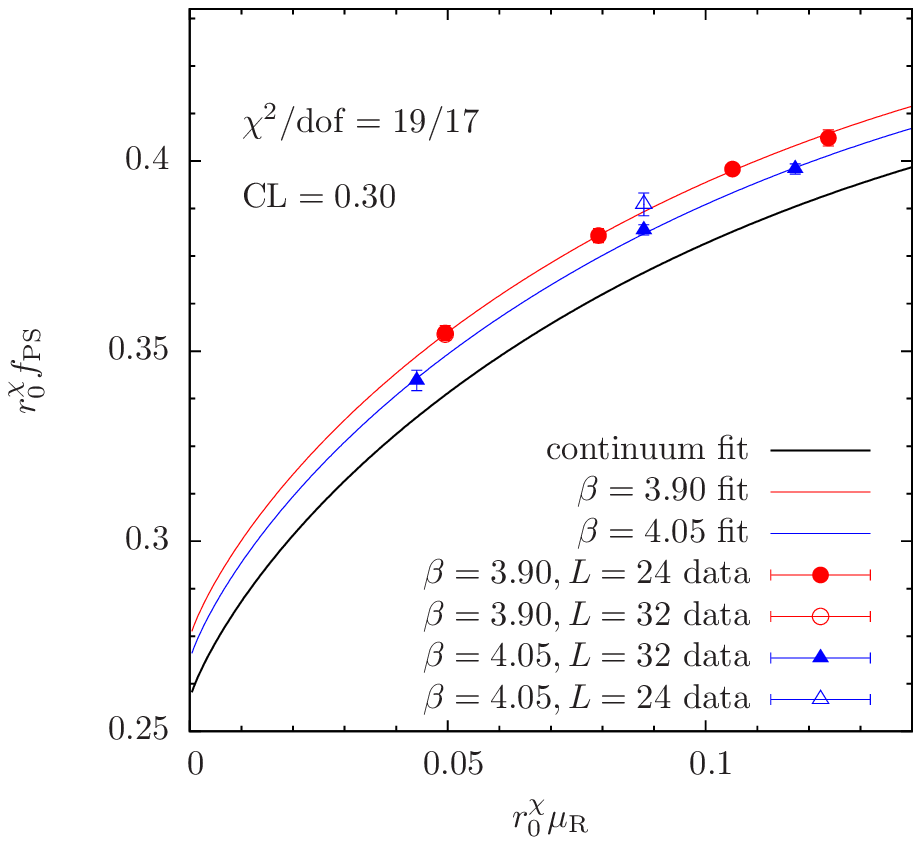}}
  \caption{Quark mass dependence:~(a) Data for $(r_0\mps)^2/r_0\mu_R$
    as a function of $r_0\mu_R$. (b) Data for $r_0\fps$ as a function
    of $r_0\mu_R$. The data are from ensembles $B_{1,2,3,4,6}$ and
    $C_{1,2,3,5}$ and the fit to this data is of type B. Note that in
    these figures we did not propagate the errors of $r_0$ and
    $Z_\mathrm{P}$.}
  \label{fig:comp}
\end{figure}

As an example we show in figures~\ref{fig:mps} and \ref{fig:fpsasq}
the result for a fit of type B on data-set composed of ensembles
$B_{1,2,3,4,6}$ and $C_{1,2,3,5}$

The main physical results we obtain from this analysis are the light
quark mass $m_{u,d}^{\overline{\mathrm{MS}}}(\mu=2\,
\mathrm{GeV})=3.54(26)\, \mathrm{MeV}$, the pseudo scalar decay
constant in the chiral limit $f_0=122(1)\, \mathrm{MeV}$, the scalar
condensate $[\Sigma^{\overline{\mathrm{MS}}}(\mu=2\,
  \mathrm{GeV})]^{1/3}=270(7)\, \mathrm{MeV}$ and
$f_\pi/f_0=1.0755(94)$. We furthermore extract accurate values for
other low energy constants of $\chi$PT, in particular
$\bar{\ell}_3=3.50(31)$ and $\bar{\ell}_4=4.66(33)$. The errors are
statistical and systematic errors summed in quadrature.

\subsection{Discussion and Conclusion}

Here we collect a short list of observations coming from a set of
$\chi$PT fits. A complete description of a large set of combined fits,
including the details on the estimates of the systematic effects, was
presented in Ref.~\cite{scalingnf2}.

We observe that including in the fits pseudoscalar masses $\mps >
520\ \mathrm{MeV}$ decreases significantly the quality of the NLO fits
($\chi^2/\mathrm{dof} \gg 1$). This indicates that the applicability
of NLO $\chi$PT in that regime of masses is disfavoured.

On the contrary, extending the fit-range to a value of $\mps\sim
280\ \mathrm{MeV}$ preserves the good quality of the fit and gives
compatible values for the fit parameters. This result makes us
confident that the extrapolation to the physical point is trustworthy.

Including lattice artifacts in the fits gives results which are
compatible to those where $D_{\mps,\fps}$ is set to zero but with a
somehow better $\chi^2/\mathrm{dof}$. We observe that the values of
the fitting parameters $D_{\mps,\fps}$ are compatible with zero within
two standard deviations. This is in line with the small discretisation
effects observed in the scaling test.

The inclusion of NNLO terms produces similar results to the NLO fits
in the quark mass region corresponding to $\mps \in
[280,520]$\,MeV. When fitting data only in this mass region ({\em
  i.e.}  when excluding from the fit the heavier masses at $\mps \sim
650$\,MeV), we observe that the fit curve at NLO lies closer to those
data points (heavier masses) than the NNLO one. On the other hand,
when including the heavier masses, the NNLO fit is able describe these
data points but the quality of the fit is somehow reduced with respect
to the NLO fit. To improve the sensitivity of our lattice data to
$\chi$PT at NNLO, additional data points would be needed.

We have presented determinations of $\fps$ and $\mps$ and their
continuum, infinite volume and chiral extrapolations. A complete
description of the results was presented in Ref.~\cite{scalingnf2}.

We thank all members of ETMC for the most enjoyable collaboration.

\bibliographystyle{h-physrev5}
\bibliography{bibliography}

\begin{thebibliography}{10}

\bibitem{Jansen:2008vs}
  K.~Jansen,
  \newblock PoS {\bf LATTICE2008}, 010 (2008), 
  \href{http://arxiv.org/abs/0810.5634}{{\tt arXiv:0810.5634 [hep-lat]}}.

\bibitem{Scholz:2009yz}
E.~E. Scholz,
\newblock \href{http://arxiv.org/abs/0911.2191}{{\tt arXiv:0911.2191
  [hep-lat]}}.

\bibitem{Boucaud:2007uk}
{\bf ETM} Collaboration, Ph.~Boucaud {\em et~al.},
\newblock Phys. Lett. {\bf B650}, 304 (2007),
  \href{http://arxiv.org/abs/hep-lat/0701012}{{\tt arXiv:hep-lat/0701012}}.

\bibitem{Urbach:2007rt}
{\bf ETM} Collaboration, C.~Urbach,
\newblock PoS {\bf LAT2007}, 022 (2007),
  \href{http://arxiv.org/abs/0710.1517}{{\tt arXiv:0710.1517 [hep-lat]}}.

\bibitem{Dimopoulos:2007qy}
{\bf ETM} Collaboration, P.~Dimopoulos, R.~Frezzotti, G.~Herdoiza, C.~Urbach
  and U.~Wenger,
\newblock PoS {\bf LAT2007}, 102 (2007),
  \href{http://arxiv.org/abs/0710.2498}{{\tt arXiv:0710.2498 [hep-lat]}}.

\bibitem{Boucaud:2008xu}
{\bf ETM} Collaboration, Ph.~Boucaud {\em et~al.},
\newblock \href{http://arxiv.org/abs/0803.0224}{{\tt arXiv:0803.0224
  [hep-lat]}}.

\bibitem{Dimopoulos:2008sy}
  {\bf ETM} Collaboration, P.~Dimopoulos, R.~Frezzotti, G.~Herdoiza, K.~Jansen, C.~Michael and C.~Urbach,
  \newblock \href{http://arxiv.org/abs/0810.2873}{{\tt arXiv:0810.2873 [hep-lat]}}.

\bibitem{scalingnf2}
  {\bf ETM} Collaboration, R.~Baron {\it et al.},
  \newblock 
  \href{http://arxiv.org/abs/hep-lat/0911.5061}{{\tt arXiv:0911.5061 [hep-lat]}}.

\bibitem{Baron:2008xa}
  {\bf ETM} Collaboration, R.~Baron {\it et al.},
  \newblock PoS {\bf LATTICE2008} (2008) 094,
  \href{http://arxiv.org/abs/hep-lat/0810.3807}{{\tt arXiv:0810.3807 [hep-lat]}}.

\bibitem{Baron:2009zq}
  {\bf ETM} Collaboration, R.~Baron {\it et al.},
  \newblock PoS {\bf LATT2009} (2009) 104,
  \href{http://arxiv.org/abs/hep-lat/0911.5244}{{\tt arXiv:0911.5244 [hep-lat]}}.

\bibitem{Weisz:1982zw}
P.~Weisz,
\newblock Nucl. Phys. {\bf B212}, 1 (1983).

\bibitem{Frezzotti:2000nk}
{\bf ALPHA} Collaboration, R.~Frezzotti, P.~A. Grassi, S.~Sint and P.~Weisz,
\newblock JHEP {\bf 08}, 058 (2001),
  \href{http://arxiv.org/abs/hep-lat/0101001}{{\tt hep-lat/0101001}}.

\bibitem{Frezzotti:2003ni}
R.~Frezzotti and G.~C. Rossi,
\newblock JHEP {\bf 08}, 007 (2004),
  \href{http://arxiv.org/abs/hep-lat/0306014}{{\tt hep-lat/0306014}}.

\bibitem{Shindler:2007vp}
A.~Shindler,
\newblock Phys. Rept. {\bf 461}, 37 (2008),
  \href{http://arxiv.org/abs/0707.4093}{{\tt arXiv:0707.4093 [hep-lat]}}.

\bibitem{Blossier:2007vv}
{\bf ETM} Collaboration, B.~Blossier {\em et~al.},
\newblock JHEP {\bf 04}, 020 (2008), \href{http://arxiv.org/abs/0709.4574}{{\tt
  arXiv:0709.4574 [hep-lat]}}.

\bibitem{Frezzotti:2008dr}
  {\bf ETM} Collaboration, R.~Frezzotti, V.~Lubicz and S.~Simula,
  \newblock Phys. Rev. {\bf D79}, 074506 (2009), \href{http://arxiv.org/abs/0812.4042}{{\tt arXiv:0812.4042 [hep-lat]}}.


\bibitem{Alexandrou:2008tn}
{\bf ETM} Collaboration, C.~Alexandrou {\em et~al.},
\newblock Phys. Rev. {\bf D78}, 014509 (2008),
  \href{http://arxiv.org/abs/0803.3190}{{\tt arXiv:0803.3190 [hep-lat]}}.

\bibitem{Jansen:2008wv}
{\bf ETM} Collaboration, K.~Jansen, C.~Michael and C.~Urbach,
\newblock \href{http://arxiv.org/abs/0804.3871}{{\tt arXiv:0804.3871
  [hep-lat]}}.


\bibitem{McNeile:2009mx}
  {\bf ETM} Collaboration, C.~McNeile, C.~Michael and C.~Urbach,
 \newblock \href{http://arxiv.org/abs/0902.3897}{{\tt arXiv:0902.3897 [hep-lat]}}.

\bibitem{Blossier:2009bx}
  {\bf ETM} Collaboration, B.~Blossier {\em et~al.},
  \newblock \href{http://arxiv.org/abs/0904.0954}{{\tt arXiv:0904.0954 [hep-lat]}}.

\bibitem{Frezzotti:2007qv}
R.~Frezzotti and G.~Rossi,
\newblock PoS {\bf LAT2007}, 277 (2007),
  \href{http://arxiv.org/abs/0710.2492}{{\tt arXiv:0710.2492 [hep-lat]}}.

\bibitem{Dimopoulos:2009qv}
P.~Dimopoulos, R.~Frezzotti, C.~Michael, G.~C. Rossi and C.~Urbach,
\newblock \href{http://arxiv.org/abs/0908.0451}{{\tt arXiv:0908.0451
  [hep-lat]}}.

\bibitem{Sommer:1993ce}
R.~Sommer,
\newblock Nucl. Phys. {\bf B411}, 839 (1994), [hep-lat/9310022].

\bibitem{Gasser:1986vb}
J.~Gasser and H.~Leutwyler,
\newblock Phys. Lett. {\bf B184}, 83 (1987).


\bibitem{Dimopoulos:2007fn}
{\bf ETM} Collaboration, P.~Dimopoulos {\em et~al.},
\newblock PoS {\bf LAT2007}, 241 (2007),
  \href{http://arxiv.org/abs/0710.0975}{{\tt arXiv:0710.0975 [hep-lat]}}.

\bibitem{Leutwyler:2000hx}
H.~Leutwyler,
\newblock Nucl. Phys. Proc. Suppl. {\bf 94}, 108 (2001),
  \href{http://arxiv.org/abs/hep-ph/0011049}{{\tt arXiv:hep-ph/0011049}}.

\bibitem{Colangelo:2005gd}
G.~Colangelo, S.~D{\"u}rr and C.~Haefeli,
\newblock Nucl. Phys. {\bf B721}, 136 (2005),
  \href{http://arxiv.org/abs/hep-lat/0503014}{{\tt hep-lat/0503014}}.

\end{thebibliography}

\end{document}